\documentclass[aps,pra,twocolumn]{revtex4-1}
\usepackage{amsmath}
\usepackage{amssymb}
\usepackage{graphics}
\usepackage{empheq}
\usepackage{ctable}
\usepackage{bbm}
\usepackage{booktabs}
\usepackage{dcolumn}
	\newcolumntype{d}[1]{D{.}{.}{#1}}
\usepackage{braket}

\newcommand\ro{\hat\rho}
\newcommand\Ho{\hat H}

\newcommand\qo{\hat q}
\newcommand\po{\hat p}
\newcommand\phio{\hat\phi}
\newcommand\chio{\hat\chi}

\newcommand\half{\frac{1}{2}}

\begin{document}

\author{Luca Ferialdi}
\email{ferialdi@ts.infn.it}
\affiliation{Department of Physics, University of Ljubljana, Jadranska 19, SI-1000 Ljubljana, Slovenia}
\affiliation{Department of Physics, University of Trieste, Strada costiera 11, 34151 Trieste, Italy}
%\title{Role of the momentum coupling in a Gaussian open-system dynamics}   
\author{Andrea Smirne}
\email{andrea.smirne@uni-ulm.de}
\affiliation{Institute of Theoretical Physics, Universit{\"a}t Ulm, Albert-Einstein-Allee 11D-89069 Ulm, Germany}
\begin{abstract}
We consider a model of non-Markovian Quantum Brownian motion that consists of an harmonic oscillator bilinearly coupled 
to a thermal bath, both via its position and momentum operators. We derive the master equation for such a model and we solve the equations of motion for a generic Gaussian system state. 
We then investigate the resulting evolution of the first and second moments for both an Ohmic and a super-Ohmic spectral density.
In particular, 
we show that, irrespective of the specific form of the spectral density, the coupling with the momentum
enhances the dissipation experienced by the system, accelerating its relaxation to the equilibrium,
as well as modifying the asymptotic state of the dynamics. Eventually, we characterize explicitly the non-Markovianity
of the evolution, using a general criterion which relies on the positivity of the master equation coefficients.
\end{abstract}
%\pacs{03.65.Yz,03.65.Ta,42.50.Lc}
\title{Momentum coupling in non-Markovian Quantum Brownian motion}   

\maketitle

\section{Introduction}
Every quantum device unavoidably interacts with the surrounding environment, which affects its dynamics. In general, such open systems are described by non-Markovian dynamics, 
which account for the memory effects involved in the influence of the environment on the relevant system~\cite{BrePet02,Wei08}. 
%With the term \lq non-Markovian\rq~ one denotes all the dynamics that do not obey the Markov approximation. 
These dynamics constitute a very large class of open-system evolutions (see the recent reviews~\cite{Rivasetal14,Breetal16,DeVAlo17}), and in order to investigate them it can be thus useful to focus on specific models.
A commonly used model is provided by a system bilinearly coupled to a bath of harmonic oscillators~\cite{CalLeg83,Foretal88,HPZ}. 
This model is at the same time physically meaningful and mathematically treatable in detail. One of the most important results of this model is the so called non-Markovian Brownian motion~\cite{HPZ,HalYu96,ForOco01}, where one considers an harmonic oscillator bilinearly coupled to a thermal bath via its position. In their seminal paper~\cite{HPZ} 
Hu, Paz and Zhang provided the exact master equation for the non-Markovian Brownian motion and analyzed its properties. 
%Other open systems (both infinite and finite dimensional) have been investigated both with approximate analytical techniques[...] and numerical methods [...].

Thanks to a recent paper~\cite{Fer16}, exact results have been extended to a wider class of systems,  including a more general form of the coupling
between the system and the environment; interestingly, the same analytical approach provides approximate results for finite dimensional systems~\cite{Fer17}.
Aim of this paper is to exploit these results to take a step forward in the understanding of non-Markovian dynamics, by investigating a non-standard model for non-Markovian Brownian motion. 
We consider an harmonic oscillator bilinearly coupled to a thermal bath, both via its position and momentum. Since the non-Markovian behavior is strictly connected to memory features of the bath, it is interesting to understand how a \lq\lq dynamical\rq\rq~system-bath coupling affects the dynamics of the system. 
In particular, we compare this extended model with the standard non-Markovian quantum Brownian motion, focusing on the new features of the dynamics provided by the momentum coupling. 
We derive the master equation fixing the open-system evolution and we describe the corresponding evolutions for the position and momentum
expectation values and variances and for the position-momentum covariance; indeed, since the dynamics preserves the Gaussian form of the reduced states,
this fully characterizes the solution of the master equation for this class of states. 
Finally, we show explicitly the non-Markovian nature of the dynamics, using the criterion for open quantum system dynamics
introduced in \cite{Hall2014}.

Let us mention that the coupling with the system's momentum has been considered in phenomenological models based on Lindblad equations~\cite{SanScu87,Gaoeco}, and stochastic Schr\"odinger equations (both for Markovian~\cite{BasIppVac05} and non-Markovian systems~\cite{FerBas12}).  
Moreover, the dissipative effects due to the coupling with the momentum instead of position (the so-called 'anomalous dissipation')
has been investigated within the context of tunneling in \cite{Leggett1984},
while the resulting thermodynamical properties has been treated in \cite{Cuccolietal2001}; eventually,
the coupling of both the system position and momentum to the bath has been considered in \cite{Kohleretal2004}
to characterize the dynamics of the relative phase in a Josephson junction, including both the fluctuations of the radiation field
and the quasiparticle tunneling.
These models indeed provide some significant examples of specific physical systems, to which the analysis of the present paper
may be applied.

The rest of the paper is organized as follows: in Sec.\ref{sec:mod} we introduce the model, we derive the exact master equation, and the evolution of relevant physical quantities. In Sec.\ref{sec:mesol} we provide a detailed analysis of the model under study for different spectral densities, and we compare its features with the standard Quantum Brownian motion. 
In Sec.\ref{sec:nm} we write down the semigroup limit of the dynamics for a $\delta$-like correlation function of the bath and discuss
the non-Markovian nature of the dynamics in the other cases.
In Sec.\ref{sec:con} we draw the conclusions.

\section{The model and its solution}\label{sec:mod}
We investigate the dynamics of a harmonic oscillator bilinearly coupled to a bosonic thermal bath via 
a linear combination of its position and momentum operators, as described by the total Hamiltonian
$\hat{H} = \Ho_S+\Ho_I+\Ho_E$, with
\begin{eqnarray}
\label{eq:ho}\Ho_S&=&\frac{\po^2}{2m}+\half m\omega_S^2\qo^2\\
\label{HI}\Ho_I&=&(\qo-\mu\po)\sum_k c_k\qo_k\\
\label{eq:hoe}\Ho_E&=&\sum_k\frac{\po_k^2}{2m_k}+\half\omega_k^2\qo_k^2,
\end{eqnarray}
where $\omega_S$ is the free frequency of the harmonic oscillator, $m$ its mass,
while $\omega_k$ and $m_k$ are, respectively, the frequency and mass of the $k$-th bath mode;
indeed, $\qo$ and $\po$ ($\qo_k$ and $\po_k$) are the system ($k$-th bath mode) position and momentum operators.
Furthermore, $\mu$ is the parameter providing us with the relative strength of the coupling with the system momentum with respect to the coupling
with the system position; as said, the effects induced by a non-zero value of the coupling $\mu$ will be one of the main focuses of our following analysis.
The bath is assumed to have a Gaussian (thermal) initial state
\begin{equation}
\rho_E(0) = \frac{e^{-\beta \hat{H}_E}}{Z}, \quad Z = \mbox{Tr}_E\left[e^{-\beta \hat{H}_E}\right]
\end{equation}
 and its action on the open system is completely characterized by the spectral density
\begin{equation}
J(\omega)=\sum_k\frac{c_k^2}{2m_k\omega_k}\delta(\omega-\omega_k)
\end{equation}
or, equivalently, by the two-point correlation function \cite{BrePet02}
\begin{eqnarray}
D(t-s)&=&\hbar\!\int_0^\infty \!\!d\omega J(\omega) \bigg[\coth\!\left(\frac{\hbar\omega}{2k_B T}\right)\!\cos\omega(t-s)\nonumber\\
&&\hspace{2.2cm}-i\sin\omega(t-s)\bigg]\,.
\end{eqnarray}

%renormalized frequency that accounts for translational invariance of the global Hamiltonian and for its boundedness [vuoi citare bassano o mettiamo caldeira-leggett?].
Before presenting the master equation and its solution for the model, let us note that with the canonical change of variables $(\hat{q},\hat{p}) \mapsto (\hat{x} = \hat{q}-\mu \hat{p},\hat{p})$, one
can equivalently describe the equations of motion 
%the equations of motion fixed by the above Hamiltonian are equivalent to those given by 
using the global Hamiltonian with the same $\Ho_E$, but where only the system operator $\hat{x}$
is coupled to the bath operator $\sum_k c_k\qo_k$, while the system free Hamiltonian is given by
\begin{eqnarray}
\Ho'_S&=&\frac{\po^2}{2 m'}+V(\hat{x},\po) 
\end{eqnarray}
with 
\begin{eqnarray}
m'&=&\left(\frac{1}{m}+ m\omega_S^2\mu^2\right)^{-1} \nonumber\\
V(\hat{x},\po) &=& \frac{m \omega_S^2}{2}\hat{x}^2+\frac{m\omega_S^2\mu}{2}(\hat{x}\po+\po\hat{x}).
\end{eqnarray}
We stress that, although one can recover a position-position coupling by means of a unitary transformation, 
the system we consider here is fundamentally different from the standard quantum Brownian motion \cite{HPZ}, the difference being now enclosed in a momentum dependent
free Hamiltonian of the system. 

\subsection{Master equation}
It has been recently shown~\cite{DioFer14,Fer16} that the exact master equation for the model
fixed by the total Hamiltonian given by Eqs.\eqref{eq:ho}-\eqref{eq:hoe}, obtained after averaging out the environmental degrees of freedom, reads
\begin{eqnarray}\label{NMdissME}
\frac{d\ro}{dt}&=&\!-\frac{i}{\hbar}[\Ho(t),\ro]+i\Xi_\mu(t)[\qo^2,\ro]+i\Upsilon_\mu(t)[\qo,\{\po,\ro\}]\\
&&+ \Gamma_\mu(t)[\qo,[\qo,\ro]]\! +\!\Theta_\mu(t) [\qo,[\po,\ro]]\! +\!\gamma_\mu(t)[\po,[\po,\ro]]\,. \nonumber
\end{eqnarray}
The second term of the right hand side yields a bath-induced frequency renormalization of the oscillator, while the third term is a dissipative contribution since it is responsible for damping of the momentum expectation value. The terms displayed by the second line of Eq.~\eqref{NMdissME} describe diffusion both in position and momentum.
The subscript $\mu$ denotes the fact that we are considering the unusual coupling \eqref{HI}.
We remark that Eq.\eqref{NMdissME} provides us with the most general form of a time-local generator such that, at any time,
the operators in the dissipator are linear in the position and momentum operators, while the Hamiltonian term
is at most quadratic with respect to them~\cite{SanScu87}.
Since the expressions of the functions displayed by the master equations (\ref{NMdissME}) as provided in~\cite{Fer16} have rather complicated expressions, it is useful to re-derive them in a more convenient way. We do so by solving the Heisenberg equations of motion of the system, by exploiting the Laplace transform $\mathcal{L}$. By introducing 
the shifted system frequency \cite{HPZ}
\begin{equation}\label{eq:omegar}
\omega_R = \sqrt{\omega_S^2+\frac{2}{m} \int d \omega \frac{J(\omega)}{\omega}}
\end{equation}
and
\begin{eqnarray}
\tilde{D}(l)&=&\mathcal{L}[D^{\mathrm{Im}}(t)]\\
G(t)&=&\mathcal{L}^{-1}\left[\frac{1}{l^2+\omega_R^2+\frac{2}{m'}\tilde{D}(l)}\right] \label{eq:gdit}
\end{eqnarray}
one finds that the solution of the equations of motion reads
\begin{eqnarray}
\label{solq}\qo(t)\!&=&\!G_1(t)\qo(0)+G_2(t)\po(0)-\!\!\int_0^t\!\!G_3(t-s)\phio(s)ds\\
\label{solp}\po(t)\!&=&\!G_4(t)\qo(0)+G_5(t)\po(0)-\!\!\int_0^t\!\!G_6(t-s)\phio(s)ds,
\end{eqnarray}
where $\phio(t)$ denotes the bath coupling operator freely evolved at time $t$:
\begin{eqnarray}
\phio(t)&\equiv&\sum_k c_k \qo_k(t) \nonumber\\
&=&\sum_kc_k\left(\qo_k(0)\cos\omega_kt+\frac{\po_k(0)}{m_k}\sin\omega_kt\right)
\end{eqnarray}
 and the Green's functions $G_i$ read
\begin{eqnarray}
G_1(t)&=&\dot{G}(t)-2\mu\int_0^tD^{\mathrm{Im}}(t-s)G(s)ds \nonumber\\
G_2(t)&=&\frac{G(t)}{m}+2\mu^2\int_0^tD^{\mathrm{Im}}(t-s)G(s)ds \nonumber\\
G_3(t)&=&\frac{G(t)}{m}+\mu \dot{G}(t) \nonumber\\
G_4(t)&=&-m\omega^2G(t)-2\int_0^tD^{\mathrm{Im}}(t-s)G(s)ds \nonumber\\
G_5(t)&=&\dot{G}(t)+2\mu\int_0^tD^{\mathrm{Im}}(t-s)G(s)ds \nonumber\\
G_6(t)&=&-m\omega^2\mu G(t)+\dot{G}(t).\label{eq:green}
\end{eqnarray}
In order to derive the master equation~\eqref{NMdissME} and its coefficients, it is convenient to introduce the characteristic operator
\begin{equation}
\chio(t)=\mathrm{Tr}_E\left[e^{i\lambda\qo(t)+i\gamma\po(t)}\ro_E(0)\right]\,.
\end{equation}
We now adopt the strategy outlined in~\cite{CarBas16}: we differentiate $\chio(t)$ with respect to $t$, and we replace the terms of the type $\qo\chio(t)$ and $\po\chio(t)$ by suitable combinations of $d\chio(t)/d\lambda$ and $d\chio(t)/d\gamma$. The equation obtained is rewritten in terms of $\chio(0)$ by exploiting the composition property of the adjoint map for $\chio$. 
After some manipulations we are able to express $d\chio(t)/dt$ in terms of (anti-)commutators of $\qo$ and $\po$ with $\chio(0)$. We exploit the following relation
\begin{equation}
\mathrm{Tr}_S\left[\frac{d\chio(t)}{dt}\ro(0)\right]=\mathrm{Tr}_S\left[\chio(0)\frac{d\ro(t)}{dt}\right]\,,
\end{equation}
and after some lengthy calculations, this procedure eventually provides us with Eq.~\eqref{NMdissME} with
\begin{eqnarray}\label{coeffME}
\Ho(t)&=&\Ho_S+\frac{\hbar\mu}{2m}\frac{H_1(t)}{F(t)} \po^2\nonumber\\
&&+\frac{\hbar}{2}\left(m\omega^2\mu^2\frac{H_1(t)}{F(t)}+\mu\frac{H_2(t)}{F(t)}\right)\{\qo,\po\} \nonumber\\
\Gamma_{\mu}(t)&=&\frac{\dot{g}_2(t)}{\hbar^2}-K_4(t)\frac{g_3(t)}{\hbar^2}-2K_2(t)\frac{g_2(t)}{\hbar^2} \nonumber\\
\label{theta}\Theta_{\mu}(t)&=&-\frac{\dot{g}_3(t)}{\hbar^2}+2K_1(t)\frac{g_2(t)}{\hbar^2}+K_5(t)\frac{g_3(t)}{\hbar^2}\\
\Xi_{\mu}(t)&=&\frac{1}{2}\frac{H_2(t)}{F(t)} \nonumber\\
\Upsilon_{\mu}(t)&=& K_5(t) \nonumber\\
\gamma_{\mu}(t)&=&\frac{\dot{g}_1(t)}{\hbar^2}-K_1(t)\frac{g_3(t)}{\hbar^2}-2K_3(t)\frac{g_1(t)}{\hbar^2}\nonumber
\end{eqnarray}
The explicit expressions for the functions displayed by these equations are provided in the Appendix A. We stress that the expressions for these functions are exact, and that when $\mu=0$ they recover those for non-Markovian Brownian motion~\cite{HPZ,HalYu96,CarBas16}, as expected.

\subsection{Time evolution of the position and momentum first and second moments}
The advantage of having solved the equations of motion in the Heisenberg picture is that they easily allow us to compute 
the expected values of relevant operators. The expectation values for $\qo$ and $\po$ follow straightforwardly from Eqs.~\eqref{solq}-\eqref{solp}, by observing that the expectation of $\phio$ is null:
\begin{eqnarray}
q_a(t)&=& G_1(t)q_a+G_2(t)p_a \nonumber\\
p_a(t)&=& G_4(t)q_a+G_5(t)p_a, \label{sigmap}
\end{eqnarray}
where we defined $q_a(t) \equiv  \mbox{Tr}\left[\qo(t) \rho\right]$ and $p_a(t) \equiv  \mbox{Tr}\left[\po(t) \rho\right]$,
with $\rho$ initial state of the system (the initial time argument will be implied from now on).
The evolution of the position variance, $\sigma_{q^2}(t) \equiv \mbox{Tr}\left[\qo(t)^2 \rho\right]-q_a(t)^2$, is obtained by squaring Eq.~\eqref{solq} and taking the expectation value, and
similarly for the momentum variance $\sigma_{p^2}(t) \equiv \mbox{Tr}\left[\po(t)^2 \rho\right]-p_a(t)^2$ 
and the position-momentum covariance $\sigma_{qp}(t) \equiv \mbox{Tr}\left[\left\{\qo(t),\po(t)\right\} \rho\right]/2 - q_a(t) p_a(t)$. In conclusion, one has
that the elements of the covariance matrix are given by
\begin{eqnarray}
\sigma_{q^2}(t)&=& G_1^2(t)\sigma_{q^2}+G_2^2(t)\sigma_{p^2}+2G_1(t)G_2(t)\sigma_{qp}-2g_1(t) \nonumber\\
\sigma_{p^2}(t)&=& G_4^2(t)\sigma_{q^2}+G_5^2(t)\sigma_{p^2}+2G_4(t)G_5(t)\sigma_{qp}-2g_2(t) \nonumber\\
\sigma_{qp}(t)&=&G_1(t)G_4(t)\sigma_{q^2}+G_2(t)G_5(t)\sigma_{p^2}\nonumber\\
&&+[G_1(t)G_5(t)+G_2(t)G_4(t)]\sigma_{qp}-g_3(t). \label{eq:covmat}
\end{eqnarray}
By virtue of these equations we can determine the position and momentum expectation values and covariance matrix at any time $t$, 
and hence any observable associated with the system's evolution, as long as one restricts
to a Gaussian initial state. Indeed, a crucial feature of the model at hand is
that the gaussianity is preserved by the dynamics, as a consequence of the bilinear structure
of the global Hamiltonian.

\section{Examples of time evolutions for an Ohmic and a super-Ohmic spectral density}\label{sec:mesol}

In this section, we provide some examples of the evolution of
the position and momentum expectation values and variances, as well as the position-momentum covariance,
focusing on the features
which trace back to the introduction of the coupling to the system's momentum, i.e. to $\mu \neq 0$.
%Since we will restrict to initial gaussian states, this will allow us to fully characterize the system statistics
%for the considered evolutions.

To get an explicit expression of the functions $G_i(t)$ and $g_i(t)$ in Eqs. (\ref{sigmap}) and (\ref{eq:covmat}), we need to specify the form
of the spectral density, which encloses the effects of the interaction with the environment on the system
dynamics. We will consider the standard case given by \cite{HPZ}
\begin{equation}\label{eq:jomega}
J(\omega) = \frac{2 m \gamma}{\pi}\omega \left(\frac{\omega}{\Omega}\right)^{s-1} e^{-\omega^2/\Omega^2},
\end{equation}
where $\Omega$ is the cut-off frequency, $\gamma$ fixes the global coupling strength,
whereas $s$ determines the low-frequency behavior and is often referred to as Ohmicity parameter:
for $s=1$ one says that $J(\omega)$ in an ohmic spectral density, while for $s>1$ ($s<1$)
one speaks about super-ohmic (sub-ohmic) spectral density.
%Finally, we use a gaussian form of the cut-off to be consistent with~\cite{HPZ}.

\subsection{Ohmic spectral density}

We start by taking into account the Ohmic case, i.e., $s=1$. 
This spectral density is known to provide the semigroup description of the open system dynamics,
in the infinite temperature and infinite cut-off limits~\cite{CalLeg83,BrePet02,Fer17b}, and then it provides us with a natural reference case.
Note that the mentioned semigroup limit is obtained also for $\mu\neq0$, as stated in~\cite{Fer17b}
and explicitly shown later on.

\begin{figure}[ht!]
{\bf (a)}\hskip4cm{\bf (b)}\\
\includegraphics[width=.5\columnwidth]{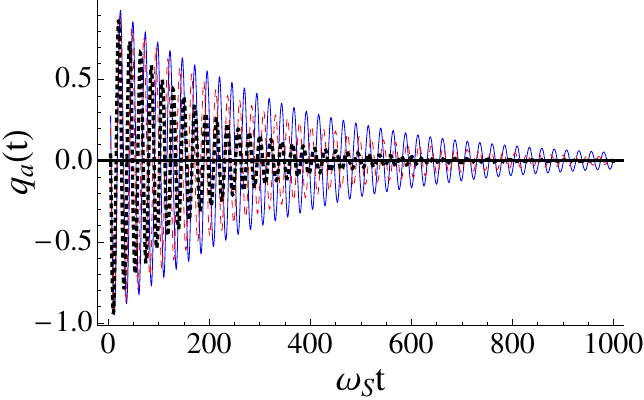}\includegraphics[width=.5\columnwidth]{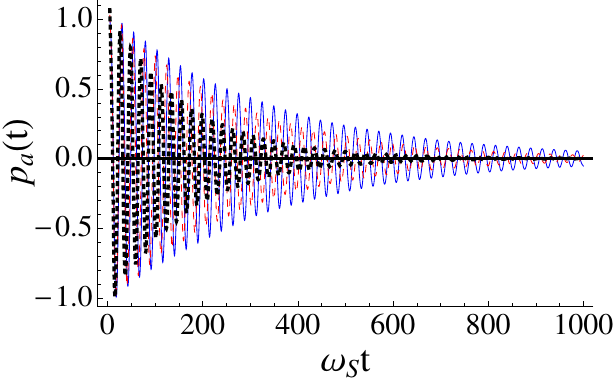}
\caption{Evolution in time of the expectation value of position {\bf(a)} and momentum {\bf(b)} [see Eq.\eqref{sigmap}], under an Ohmic spectral density, $s=1$ in Eq.\eqref{eq:jomega}.
The different lines correspond to different values of the coupling strength with the system momentum,
$\mu=0$ (blue solid line), $m\mu \omega_S=0.5$ (red dashed line), $m \mu \omega_S=1$ (black dotted line); the other parameters are
$\gamma/\omega_S =3*10^{-3} $, $\Omega/\omega_S = 20$ and $\hbar \beta \omega_S = 10^{-2}$, while as initial conditions
we set $\sqrt{m \omega_S/\hbar}\, q_a =1$,  $p_a/\sqrt{m \omega_S \hbar} = 10^{-2}$ and $ (m \omega_S/\hbar)\sigma_{q^2} = 0.5$; 
the expectation values of position and momentum are expressed in units of, respectively, $\sqrt{\hbar/(m \omega_S)}$ 
and $\sqrt{m \omega_S \hbar}$.}
\label{fig:1}
\end{figure}

First, in Fig.\ref{fig:1}.{\bf (a)} and {\bf (b)} we see the time evolution of the expectation
values of, respectively, position and momentum for different values of the coupling parameter $\mu$.
In both cases, and for any value of $\mu$, we have decaying oscillations to the asymptotic value zero.
On the other hand, the introduction of a coupling with the system momentum
accelerates the relaxation process of both the quantities, which is the faster the higher the value of $\mu$.
The coupling with the momentum brings along a further contribution to the friction experienced by the open system
due to its coupling with the environment, so that the damping of the momentum itself is enhanced. 
Indeed, referring to the master equation (\ref{NMdissME}), it is clear how this phenomenon can be traced back to the changes in the friction coefficient $\Upsilon_\mu(t)$, 
which now depends on the coupling $\mu$ (all the other terms vanish when one takes the expectation value with the momentum operator).

\begin{figure}[ht!]
{\bf (a)}\hskip4.5cm{\bf (b)}\\
\vspace{0.2cm}
\includegraphics[width=.48\columnwidth]{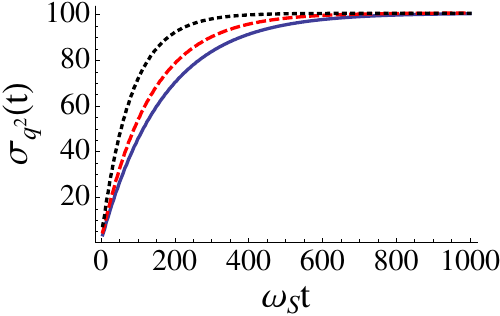}
\includegraphics[width=.48\columnwidth]{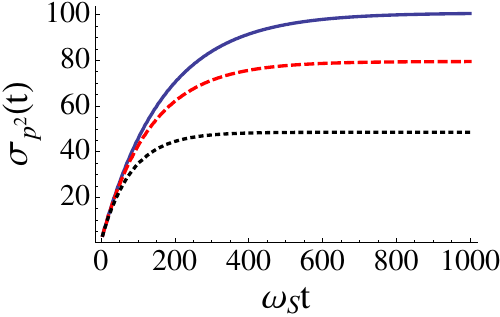}\\
{\bf (c)}\\
\vspace{0.3cm}
\includegraphics[width=.58\columnwidth]{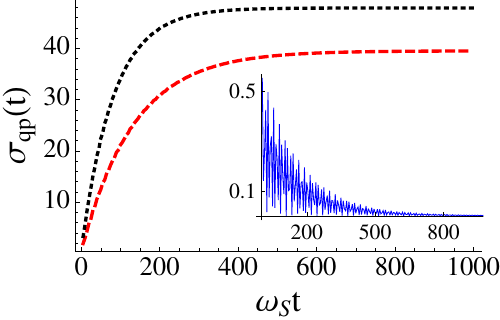}
\caption{Evolution in time of the elements of the covariance matrix, see Eq.\eqref{eq:covmat}, under an Ohmic spectral density, $s=1$ in Eq.\eqref{eq:jomega}:
variance of the position $\sigma_{q^2}(t)$ in {\bf(a)},  variance of the momentum $\sigma_{p^2}(t)$ in {\bf(b)}
and position-momentum covariance $\sigma_{qp}(t)$ in {\bf(c)}.
The different lines correspond to $\mu=0$ (blue solid line), $m\mu \omega_S=0.5$ (red dashed line), $m\mu \omega_S=1$ (black dotted line). The other parameters are as in Fig.\ref{fig:1};
the position variance is expressed in units of $\hbar/(m \omega_S)$, the momentum variance in units of $m \omega_S \hbar$
and the position-momentum covariance in units of  $\hbar$.
The inset in {\bf(c)} magnifies the case of $\mu=0$.}
\label{fig:2}
\end{figure}

Now, let us move our numerical analysis to the elements of the system covariance matrix, which, as said,
completes the description of the reduced observables if we restrict to Gaussian states.
In Fig.\ref{fig:2} {\bf (a)}, {\bf (b)} and {\bf (c)}, we report the evolution of, respectively, $\sigma_{q^2}(t)$, $\sigma_{p^2}(t)$ and $\sigma_{qp}(t)$
for  different values of $\mu$. Once again, we note how the relaxation toward the asymptotic value is the faster 
the higher the strength of the momentum coupling. 
However, now the asymptotic values themselves of $\sigma_{p^2}(t)$ and $\sigma_{qp}(t)$
are drastically changed by a non-zero value of $\mu$: the former is decreased, while the latter is increased. 
The asymptotic expectation value of the system
kinetic energy $\po^2/(2m)$ and, as a consequence, the asymptotic expectation value of the overall system free energy, $\Ho_S$ in Eq.\eqref{eq:ho}, is progressively decreased by an increasing value of $\mu$:
the coupling with the momentum intensifies and accelerates the dissipation of the open-system. 
In addition, the whole
evolution of $\sigma_{qp}(t)$ is qualitatively modified: we have
a (non-monotonic, see the inset) relaxation to the 0 value for $\mu=0$, while 
there is a monotonically increasing evolution to a non-zero asymptotic value for $\mu\neq 0$; note that such monotonicity can be lost for different initial conditions (see below).
The coupling with the momentum and the subsequent new terms in the master equation (\ref{NMdissME})
imply that the Gibbs state is no longer the equilibrium state of the reduced dynamics, which, instead,
exhibits a non-zero value of $\sigma_{qp}$~\cite{SanScu87}.
Overall, the introduction of $\mu \neq 0$ squeezes the momentum uncertainty of the asymptotic state
and adds a non-trivial correlation among the momentum and position statistics. 

\begin{figure}[ht]
{\bf (a)}\hskip4.5cm{\bf (b)}\\
\includegraphics[width=.5\columnwidth]{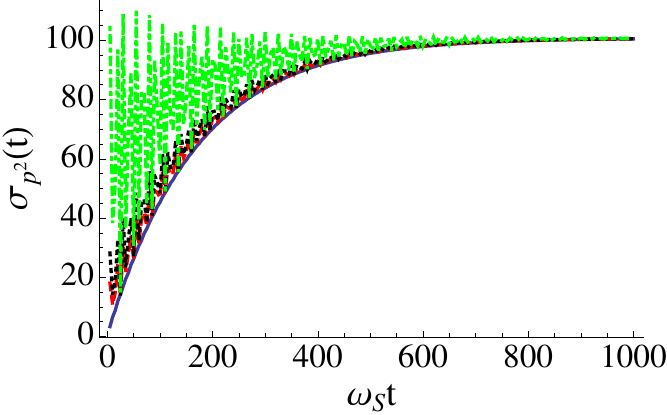}\includegraphics[width=.5\columnwidth]{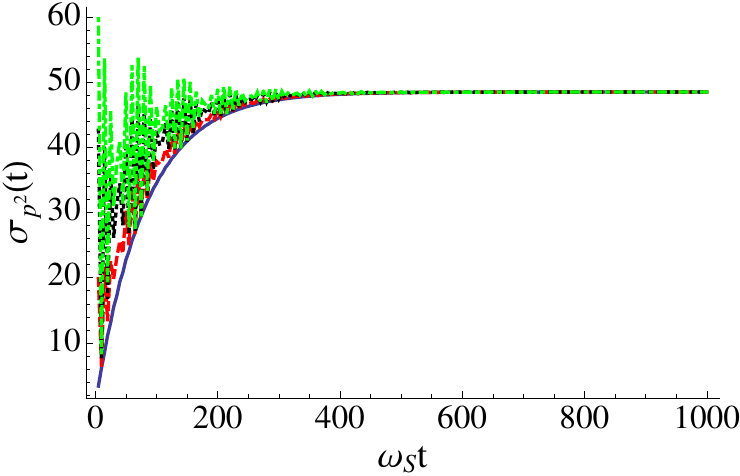}\\
{\bf (c)}\hskip4.5cm{\bf (d)}\\
\vspace{0.2cm}
\includegraphics[width=.5\columnwidth]{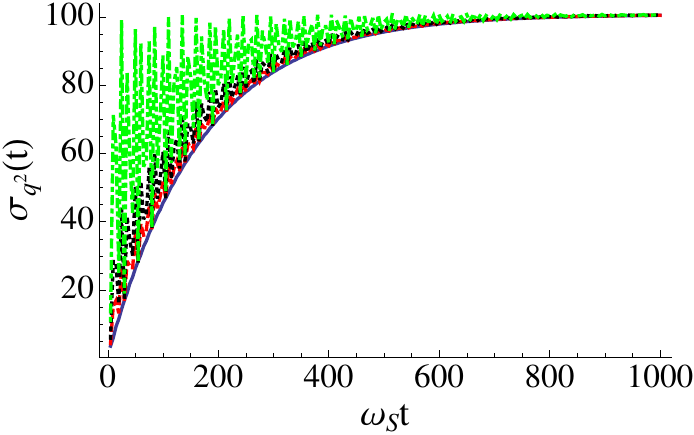}\includegraphics[width=.5\columnwidth]{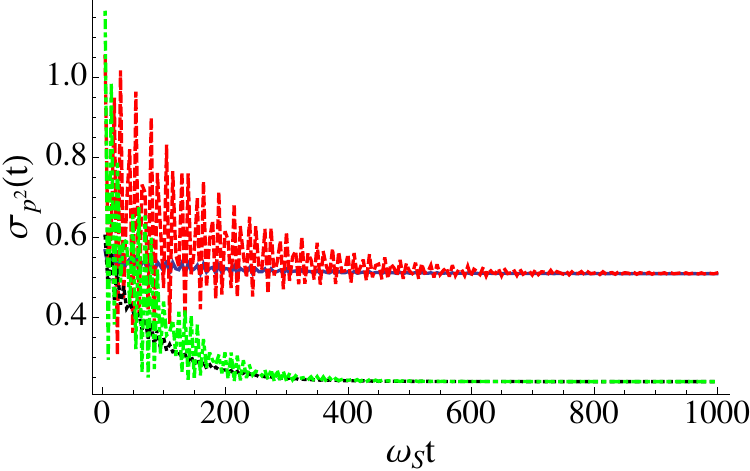}
\caption{Relaxation to the equilibrium of the momentum variance for $\mu = 0$,  {\bf(a)}, and $m\mu\omega_S=1$,  {\bf(b)},
and relaxation to the equilibrium of the position variance for $\mu =0$,  {\bf(c)}, for an Ohmic spectral density.
The different lines correspond to different initial gaussian states.
{\bf (d)} Relaxation to the equilibrium of the momentum variance at zero temperature, $T=0$, for an Ohmic
spectral density and $\mu=0$ (black, solid line and red, dashed line)
and $m\mu \omega_S=1$ (black, dotted line and green, dot-dashed line); for each value of $\mu$,
the two lines correspond to different initial gaussian states; the other parameters
are as in Fig.\ref{fig:2}.
%: $\sigma_{q^2}(0)=5*10^{-3}$, $\sigma_{p^2}(0)=50$, $x_a(0)=p_a(0)=0$ (blue, solid line), 
%$\sigma_{q^2}(0)=0.15$, $\sigma_{p^2}(0)=1.67$, $x_a(0)=0$, $p_a(0)=120$ (red, dashed line), 
%$\sigma_{q^2}(0)=0.35$, $\sigma_{p^2}(0)=0.71$, $x_a(0)=p_a(0)=0$ (black, dotted line), 
%$\sigma_{q^2}(0)=1$, $\sigma_{p^2}(0)=0.25$, $x_a(0)=p_a(0)=0$ (green, dot-dashed line);
%$\sigma_{qp}(0)$ is always set to 0, the other parameters are as in Fig.\ref{fig:1}.
}
\label{fig:3}
\end{figure}

Until now, we have considered the evolution of the momentum and position expectation values and covariances
for a fixed initial condition. Additionally, we verified numerically that the discussed asymptotic values
do not depend on the initial conditions (at least, as long as one stays within the set of initial gaussian states).
Representative examples
are given in Fig \ref{fig:3} {\bf (a)} and {\bf (b)}
for the evolution of $\sigma_{p^2}(t)$ with, respectively, $\mu=0$ and $\mu \neq 0$ and in Fig \ref{fig:3} {\bf (c)}
for the position variance with $\mu=0$;
fully analogous results hold for the other elements of the covariance matrix and for the expectation values (for the considered values of the model parameters).
Thus, the system relaxes to a unique asymptotic state, both for $\mu=0$ and $\mu \neq 0$; indeed, as previously shown, such state
will be different in the two cases.

Moreover, from Fig \ref{fig:3} {\bf (a)}, {\bf (b)} and {\bf (c)} we can observe that, for certain initial conditions,
also the position and momentum variances relax to the asymptotic value in a non-monotonic way, as we already observed for the expectation values. 
%For some initial conditions, 
%both the momentum and the position variance show even strong oscillations during their evolution.
Each variance can show even strong oscillations when its initial value is high enough and
the oscillations are the wider the higher such initial value is.
Comparing Fig \ref{fig:3} {\bf (a)} and {\bf (b)}, one can see how the feature is present both for 
$\mu=0$ and for $\mu\neq0$. The only effect 
of the coupling to the system momentum is the appearance of some beats in the oscillating
evolutions of the variances.
%In order to gain some understanding about the origin of these oscillations, we considered how they are
%modified by taking into account a different initial temperature of the bath. 

Note that all the previous examples concern the high-T regime. Nevertheless,
the results in Eqs. \eqref{sigmap} and \eqref{eq:covmat} are referred to a completely generic temperature.
In particular, one can readily see how the evolution of the momentum and position expectation values is not affected
by a change in $T$ (since the two quantities do not depend on $D^{\text{re}}(t)$, see Eqs. \eqref{eq:gdit}, \eqref{eq:green}, and \eqref{sigmap}). On the other hand,
the temperature influences the evolution of the elements
of the covariance matrix, and, especially, their asymptotic values.
In Fig.\ref{fig:3} {\bf (d)}, we study the relaxation to the equilibrium of the momentum variance
for different initial conditions and different values of $\mu$, at $T=0$.
Of course, the zero-temperature environment makes the system's momentum variance relax
to a smaller value, compared to the high-$T$ regime, while the qualitative
behavior of the whole time-evolution is rather similar for the two temperature regimes. 
Importantly for our purposes, we note that also for $T=0$, as previously described for the high-$T$ regime, introducing a non-zero value of $\mu$
affects the relaxation process by accelerating it and changing the asymptotic values; 
Fig.\ref{fig:3} {\bf (d)} shows how the asymptotic value of $\sigma_{p^2}(t)$
for $\mu \neq 0$ is decreased, with respect to the case $\mu=0$.
Finally, we also recover that a non-zero value of $\mu$
may induce some beats in the oscillating evolution of $\sigma_{p^2}(t)$.

\subsection{Super-ohmic spectral density}

%%NNBB Cambiare direttamente il valore di gamma, riassorbendo il fattore diverso e drilo che in questo modo la forza overall  la stessa!!!
Here, we examine the behavior of the system first and second moments for a non-ohmic
spectral density, in order to show that the conclusions we drew previously about
the effects of the coupling $\mu \neq 0$ do not depend on the peculiar
case given by the Ohmic spectral density.
Besides $s$, the other parameters are the same as those of the previous paragraph, with the
exception of the coupling constant $\gamma$, which has been set
so to keep unchanged the overall strength of the coupling to the bath, as quantified by $\int d \omega J(\omega)$.
Note that also the renormalized frequency changes due to the different spectral density, see Eq.\eqref{eq:omegar}

\begin{figure}[ht]
{\bf (a)}\hskip4.5cm{\bf (b)}\\
\includegraphics[width=.5\columnwidth]{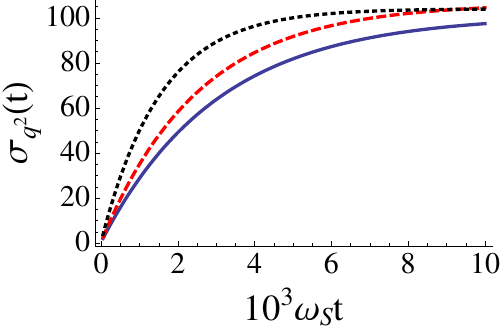}\includegraphics[width=.5\columnwidth]{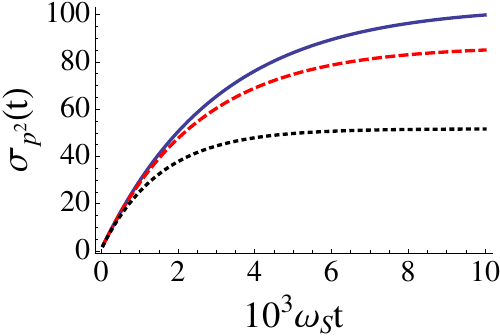}\\
{\bf (c)}\\
\includegraphics[width=.6\columnwidth]{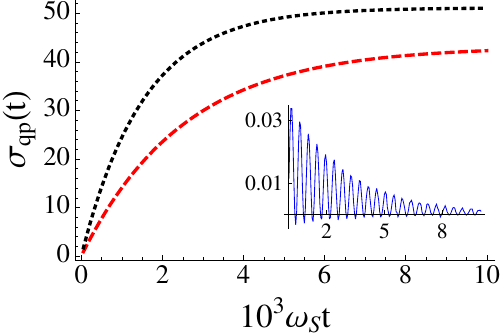}
\caption{Evolution in time of the elements of the covariance matrix, see Eq.\eqref{eq:covmat}, under a super-ohmic spectral density, $s=2$ in Eq.\eqref{eq:jomega}:
variance of the position $\sigma_{q^2}(t)$ in {\bf(a)},  variance of the momentum $\sigma_{p^2}(t)$ in {\bf(b)}
and position-momentum correlation $\sigma_{qp}(t)$ in {\bf(c)}.
The different lines correspond to $\mu=0$ (blue solid line), $m\mu\omega_S=0.5$ (red dashed line), $m\mu\omega_S=1$ (black dotted line); the other parameters are as in Fig.\ref{fig:1},
apart from $\gamma/\omega_S = 3.4*10^{-3}$.
The inset in {\bf(c)} magnifies the case of $\mu=0$.}
\label{fig:4}
\end{figure}
In particular, we considered the case $s=2$, i.e. a superohmic spectral density.
The most relevant effect due to the transition from an ohmic
to a superohmic spectral density is that the dynamics is strongly slowed down.
This can be observed from the plots in Fig.(\ref{fig:4}) [note the different
scale in the time axis compared to the plots in Fig.(\ref{fig:2})], where we reported the evolution
of the position and momentum variances and covariance; indeed, the same behavior
could be observed looking at the momentum and position expectation values.
The slowing down of the system dissipation, which is already well-known~\cite{HPZ}
in the case $\mu=0$, remains essentially unaltered, i.e., on the same timescales, also in the presence of the coupling with the system momentum.
On the other hand, one can see how a non-zero value of $\mu$ introduces some changes in the system dynamics,
which are essentially the same as for the Ohmic case.  
The relaxation process is accelerated, with respect to $\mu = 0$,
due to the further contributions to friction and dissipation: the asymptotic values are approached
in a shorter time and the asymptotic value of the system free energy is the smaller the higher $\mu$.
Moreover, as for the Ohmic case, the evolution of $\sigma_{qp}(t)$ is also qualitatively modified, leading to an asymptotic non-zero value.

The asymptotic values are
slightly increased by the super-Ohmicity of the spectral density; nevertheless,
the effects of $\mu \neq 0$ are even quantitatively very close to the Ohmic case: 
the ratio among the asymptotic values for different values of $\mu$
is approximately the same for the Ohmic and the super-Ohmic case, as shown in table I.

\begin{table}[htp]
\begin{center}
\begin{tabular}{|c|c|c|}
           Asymptotic ratio & $s=1$ & $s=2$\\
            \specialrule{.2em}{.2em}{.2em}
       $\sigma^{\infty}_{p^2}(0)/\sigma^{\infty}_{p^2}(0.5)$ & $1.20$ & $1.27$ \\[0.2cm]
       \hline
	 $\sigma^{\infty}_{p^2}(0)/\sigma^{\infty}_{p^2}(1)$ & $2.00$ & $2.07$\\[0.2cm]
	 \hline
	 $\sigma^{\infty}_{qp}(0.5)/\sigma^{\infty}_{qp}(1)$ & $0.84$ & $0.82$\\[0.2cm]
	 \hline
\end{tabular}
\end{center}
\label{default}
\caption{Ratio among the asymptotic values for the momentum variance, $\sigma^{\infty}_{p^2}(m\mu \omega_S)$, 
and the position-momentum covariance, $\sigma^{\infty}_{qp}(m\mu\omega_S)$, (the position variance does not change)
for different values of $\mu$, for the Ohmic (left column) and the super-Ohmic (right column) spectral densities.}
\end{table}%

\begin{figure}[ht]
 {\bf (a)}\hskip4.5cm{\bf (b)}\\
 \vspace{0.2cm}
\includegraphics[width=.5\columnwidth]{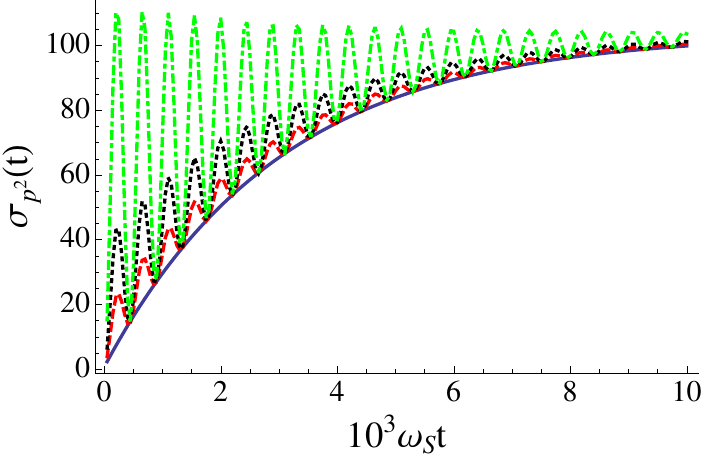}\includegraphics[width=.5\columnwidth]{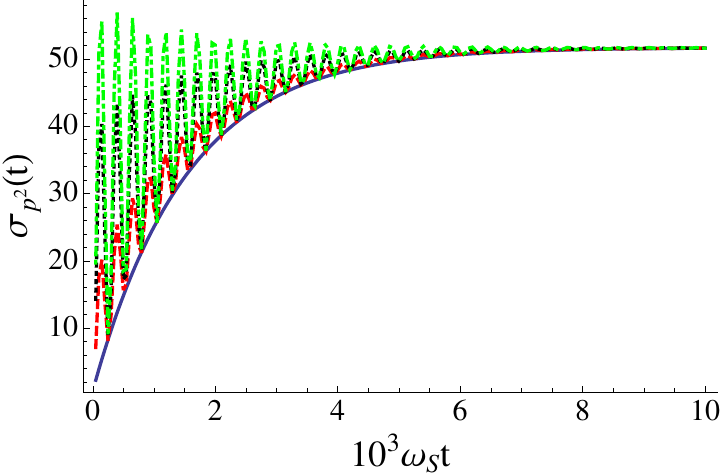}
\caption{Relaxation to the equilibrium of the momentum variance for $\mu = 0$,  {\bf(a)}, and $\mu=10^{-2}$,  {\bf(b)}, 
for a super-Ohmic spectral density with $s=2$;
the different lines correspond to different initial gaussian states. The other parameters are as in Fig.\ref{fig:5}.}
\label{fig:5}
\end{figure}

Finally, we checked also the relaxation to a unique asymptotic state, within the set of initial gaussian conditions.
In Fig.\ref{fig:5} {\bf (a)} and {\bf (b)}, we reported the evolution of the momentum variance for, respectively, $\mu=0$ and $\mu \neq 0$.
In both cases, one has a convergence to the same asymptotic value on longer time-scales.
Moreover, we note that, also in the super-Ohmic case, high enough initial values of the variance lead to an oscillating behavior, for both $\mu = 0$ and $\mu \neq 0$.
A non-zero value of $\mu$ now increases the amplitude of the oscillations for certain initial conditions, 
but without leading to the appearance of the beats as in the Ohmic case.

\section{Non-Markovianity of the dynamics}\label{sec:nm}
In this section we show explicitly that the dynamics of the model we are describing is generally non-Markovian, according to one of
the definite notions of quantum Markovianity which have been widely discussed in the literature (see \cite{Rivasetal14,Breetal16,DeVAlo17}
and references therein). In particular, we will adopt the definition which identifies quantum Markovian dynamics with those dynamics
characterized by a time-local master equation with (possibly time-dependent) positive coefficients \cite{Hall2014}.
We first briefly recall the definition for finite dimensional systems and then we apply it to the system we are dealing with here.

Hence, consider the open-system dynamics described by the one-parameter family of completely positive (CP) maps $\left\{\Lambda(t)\right\}_{t\geq0}$
and the associated time-local master equation
\begin{equation}
\frac{d}{d t} \rho(t) = \mathcal{K}(t) \rho(t);
\end{equation}
the possible presence of times where the time-local generator $\mathcal{K}(t)$ does not exist would not affect the following discussion.
Now, given a system associated with the finite dimensional Hilbert space $\mathbbm{C}^N$, 
the time-local generator $\mathcal{K}(t)$ can be always written in the form
\begin{eqnarray}\label{lin}
&&\mathcal{K}(t) \rho =-i[\hat{H},\rho]+\sum^{N^2}_{ij=1} a_{ij}(t)\left(\hat{G}_i\rho\hat{G}_j^{\dag}-\half\left\{\hat{G}_j^{\dag}\hat{G}_j,\rho\right\}\right),\nonumber\\
\end{eqnarray}
as a consequence of trace and hermiticity preservation \cite{Gorini1976}.
Here, $\hat{H}$ is an hermitian operator, $\left\{G_i\right\}_{i=1,\ldots N^2}$ is a generic basis in the set of linear operators on $\mathbbm{C}^N$
and the coefficients $a_{ij}(t)$ define an hermitian matrix, the so-called Kossakowski matrix, at any time $t$: set $(A(t))_{i j} \equiv a_{ij}(t)$,
one has $A^{\dag}(t) = A(t)$. 
Therefore, one can always diagonalize $A(t)$, via the unitary matrix $V(t)$, so that $A(t) = V(t) D(t) V^{\dag}(t)$,
with $D(t) = \mbox{diag}\left\{d_1(t), \ldots d_{N^2}(t)\right\}$ and the $d_i(t)$s are real functions of time.
As a consequence, introducing the time-dependent (Lindblad) operators $L_i(t) = \sum_{j} U_{j i}(t) G_j$,
one gets the canonical diagonal form \cite{Bengtsson2006} of the time-local generator
\begin{eqnarray}\label{lin2}
&&\mathcal{K}(t)\rho =-i[\hat{H},\rho]+\\
&&\sum^{N^2}_{i=1} d_{i}(t)\left(\hat{L}_i(t)\rho\hat{L}_i^{\dag}(t)-\half\left\{\hat{L}_i^{\dag}(t)\hat{L}_i(t),\rho\right\}\right).\nonumber
\end{eqnarray}
Now, the definition introduced in \cite{Hall2014} identifies Markovian dynamics with those dynamics where $d_i(t) \geq 0$
for any $i$ and for any time $t$. In the special case of constant positive coefficients, we thus recover the Lindblad master equation \cite{Lind1976,Gorini1976,BrePet02}, which corresponds
to the case of a Markovian time-homogeneous dynamics. 
The mentioned definition further identifies Markovian time-inhomogenous dynamics
with those given by a master equation with time-dependent positive coefficients. 
Finally, the presence of time intervals where some coefficient is negative 
is equivalent to the occurrence of a non-Markovian dynamics.
Indeed, the condition about the positivity of the coefficients of the diagonal form of the time-local generator in Eq.\eqref{lin2} can be equivalently
expressed in terms of the positive definitiveness of the Kossakowski matrix $A(t)$ in Eq.\eqref{lin}.

In order to extend the previous definition to the open-system dynamics we are studying here, which involves
a master equation for an infinite dimensional space and with unbounded operators, we can simply proceed as follows. 
We rewrite our master equation in the non-diagonal form:
\begin{equation}\label{HPZlin}
\frac{d\ro}{dt}=-i[\hat{\widetilde{H}},\ro]+\sum_{i,j} a_{ij}(t)\left(\hat{F}_i\ro\hat{F}_j-\half\left\{\hat{F}_j\hat{F}_i,\ro\right\}\right)
\end{equation}
with $\hat{F}_1=\qo$, $\hat{F}_2=\po$,
\begin{equation}\label{eq:ref}
\hat{\widetilde{H}} = \Ho(t)-\hbar\Xi_{\mu}(t)\qo^2-\frac{\hbar}{2}\Upsilon_{\mu}(t)\{\qo,\po\}
\end{equation}
 and $a_{ij}(t)$ matrix elements of
\begin{equation}\label{kossa}
A(t)=\left(
\begin{array}{cc}
-2\Gamma_{\mu}(t)&-\Theta_{\mu}(t)+i\Upsilon_{\mu}(t)\\
-\Theta_{\mu}(t)-i\Upsilon_{\mu}(t)&-2\gamma_{\mu}(t)
\end{array}\right)\,.
\end{equation}
As common in the literature we will still call this Kossakowski matrix, although it is not referred to a basis in the linear space 
of operators on the (infinite dimensional) Hilbert space associated with our system.
Now, we can identify Markovian dynamics precisely with those dynamics where the Kossakowski matrix $A(t)$ is positive definite, 
and hence the resulting diagonal time-local master equation is fixed by positive coefficients.
%Indeed, the Kossakowski matrix in Eq.\eqref{kossa}  Nevertheless, 
%it is referred to the position and momentum operators, which generate the algebra of the observables, so that its use to discriminate between Markovian and non-Markovian dynamics can be seen as a convenient generalization of the construction made in the finite dimensional case.

With this definition at hand, we first note that
if the bath correlation function is 
proportional to a Dirac delta, $D(t-s) = C \delta(t-s)$, $C>0$,
the master equation \eqref{NMdissME} reduces to
\begin{equation}\label{MdissME}
\frac{d\ro}{dt}=-\frac{i}{\hbar}[\Ho,\ro]- C [\qo,[\qo,\ro]]+2\mu C [\qo,[\po,\ro]] -\mu^2 C[\po,[\po,\ro]]\,,
\end{equation}
with $\Ho=\Ho_0-\hbar C \qo^2-\hbar C \mu^2\po^2+\hbar C \mu\{\qo,\po\}$,
which can be cast in the Lindblad form
\begin{eqnarray}
\frac{d\ro}{dt}&=&-\frac{i}{\hbar}[\Ho,\ro]+\gamma  \left( \hat{L} \ro \hat{L}-\frac{1}{2}\left\{\hat{L}^2, \ro\right\} \right);\label{MdissMEdiag}\\
\gamma &\equiv& 2 C \quad \hat{L} \equiv \left(\qo-\mu \po\right)\,. \nonumber
\end{eqnarray}
A delta-correlated (or uncorrelated) two-point function for the bath can be obtained by considering an Ohmic spectral density and taking the limits for temperature and cut-off to infinity (see e.g.~\cite{CalLeg83}); 
in this case the constant $C$ is proportional to the temperature itself, $C= 2m\gamma/\hbar^2\beta$. 
Moreover, for $\mu=0$ one recovers the Joos and Zeh master equation~\cite{JoosZeh}, as one expects from a non-dissipative Markovian dynamics (see~\cite{Fer17b} for further comments on this issue).

For all the other bath correlation functions, the
Kossakowski matrix in Eq.\eqref{kossa} is not positive definite: one of its eigenvalues
is always negative, as can be shown by evaluating the determinant of $A(t)$.
Actually, to do that it is convenient to exploit the coefficients of the master equation as derived with the method of~\cite{Fer16},
being the expressions in Eq.~\eqref{coeffME} rather involved.
In Appendix \ref{app:B} we evaluate explicitly the determinant of $A(t)$,
getting
\begin{eqnarray}
\mathrm{det}[a(t)]&\equiv&4\Gamma_{\mu}(t)\gamma_{\mu}(t)-\left(\Theta_{\mu}(t)^2+\Upsilon_{\mu}(t)^2\right)\nonumber\\
\label{deta}&=&-\left[\left(\Theta_{\mu}(t)+\mu\Gamma_{\mu}(t)\right)^2+\Upsilon_{\mu}(t)^2\right],\label{eq:at}
\end{eqnarray}
which is negative for any non-singular bath correlation function (while for a singular bath correlation function it is equal to 0, see also Eq.\eqref{MdissMEdiag}).
Accordingly, the dynamics of the system, apart from the special case of a delta-correlated bath, is always non-Markovian.
We conclude that, as was argued in \cite{Fer17b}, the master equation \eqref{NMdissME} with coefficients as in Eq.\eqref{coeffME} can describe
a time-homogeneous Markovian (i.e. semigroup) dynamics, as a singular limiting case, but it never yields a time-inhomogeneous Markovian dynamics.
Note that analogous results have been obtained in \cite{Li2017} for the spin-boson master equation derived in \cite{Fer17}. 

As a final remark, let us note that the connection between the positivity of the coefficients
of the diagonal time-local master equation and other definitions of quantum Markovianity becomes more subtle in the infinite dimensional case. 
In particular, let us mention CP-divisibility, i.e., the property of the dynamical maps 
of being, not only CP, but also decomposable intro CP terms, according to $\Lambda(t) = \Phi(t,s) \Lambda(s)$,
where $\Phi(t,s)$ are CP maps, for any $t \geq s$. This property has been identified with quantum Markovianity in \cite{Rivas2010}
and, in the finite dimensional case, one can show quite straightforwardly 
that the dynamics is CP-divisible if and only if the coefficients of the master equation \eqref{lin2} are non-negative at any time \cite{Laine2010,Rivas2012}.
Such an equivalence is not a-priori guaranteed in the infinite dimensional case, 
due to the lack of a general theorem about the generator of CP semigroups involving unbounded operators \cite{Holevo1997}.
On the other hand, in the presence of Gaussian-preserving dynamics and if one restricts to Gaussian states, CP-divisibility can be formulated by means of definite conditions, possibly expressed in terms 
of the matrices fixing the evolution of the expectation values and covariance matrix \cite{Torre2015,Liuzzo2017}. Moreover, also in infinite- dimensional systems non-Markovianity can be traced back to a nonmonotonic time evolution of proper quantities~\cite{Vasile2011,35,36}

\section{Conclusions}\label{sec:con}
We have investigated a model for non-Markovian Quantum Brownian motion where the system is bilinearly coupled to a bosonic bath, 
not only via its position, but also via its momentum.
By means of the exact master equation, along with the solution of the equations of motions for the first and second momenta
of the position and momentum operators,
we have studied the contributions to friction and dissipation induced by such an unusual momentum coupling.
The latter induces a faster relaxation to the asymptotic steady state,
characterized by a smaller average free energy, along with the appearance of a significant correlation between the position and the momentum statistics.
These results hold for different spectral densities (Ohmic and super-Ohmic), 
as well as different bath temperatures and system initial states.

In addition, we have also clarified the non-Markovian nature of the dynamics.
We have shown that the exact model at hand includes as a limiting case the time-homogeneous Markovian (i.e. semigroup) dynamics,
but it never describes a time-inhomogeneous Markovian dynamics.\\

\acknowledgments The authors wish to thank M. Carlesso for fruitful discussions. The work of LF was supported by the TALENTS$^3$ Fellowship Programme, CUP code J26D15000050009, FP code 1532453001, managed by AREA Science Park through the European Social Fund. The work of A.S. was supported by QUCHIP Project (GA No. 641039).

\appendix

\section{Analytic expressions for the functions defining the coefficients of the time-local master equation.}
In this Appendix we provide the analytic expressions for the functions displayed by Eqs.~\eqref{theta}. It is useful to introduce the \lq\lq average Green function\rq\rq~$\bar{G}$:
\begin{equation}
\bar{G}(t)=\int_0^t D(t-s) G(s)ds\,,
\end{equation}
and its suitable combinations with the Green function and its derivatives (the dot over a symbol denotes differentiation with respect to time):
\begin{eqnarray}
F(t)&=&\hbar[\dot{G}(t)\dot{G}(t)-\ddot{G}(t)G(t)]\,,\\
H_1(t)&=&\bar{G}(t)\dot{G}(t)-\dot{\bar{G}}(t)G(t)\,,\\
H_2(t)&=&\dot{\bar{G}}(t)\dot{G}(t)-\ddot{G}(t)\bar{G}(t)\,.
\end{eqnarray}
These function are the building blocks of coefficients of the master equation~\eqref{NMdissME}.
The lengthy procedure described in Sec.II.A eventually provides us with the following functions displayed by Eqs.~\eqref{theta}:
\begin{eqnarray}
K_1(t)&=&\frac{1}{m}+\frac{\hbar\mu}{m}\frac{H_1(t)}{F(t)} \nonumber\\
K_2(t)&=&\frac{\hbar}{m}\frac{H_1(t)}{F(t)}-\hbar\mu\frac{H_2(t)}{F(t)} \nonumber\\
K_3(t)&=&\hbar m\omega_S^2\mu^2\frac{H_1(t)}{F(t)}+\hbar\mu\frac{H_2(t)}{F(t)} \nonumber\\
K_4(t)&=&-m\omega_S^2+\hbar\frac{H_2(t)}{F(t)} \nonumber\\
K_5(t)&=&\left(\frac{1}{2m}+\frac{m\omega_S^2\mu^2}{2}\right)\frac{H_1(t)}{F(t)} \nonumber\\
g_1(t)&=&-\frac{1}{4}\int_0^tds\int_0^tdlD^{\mathrm{Im}}(s-l)G_3(t-s)G_3(t-l) \nonumber\\
g_2(t)&=&-\frac{1}{4}\int_0^tds\int_0^tdlD^{\mathrm{Im}}(s-l)G_6(t-s)G_6(t-l) \nonumber\\
g_1(t)&=&-\frac{1}{2}\int_0^tds\int_0^tdlD^{\mathrm{Im}}(s-l)G_3(t-s)G_6(t-l) \nonumber\,.
\end{eqnarray}
\\

\section{Derivation of Eq.\eqref{eq:at}.}\label{app:B}

The elements of the Kossakovski matrix~\eqref{kossa} as derived with the technique of~\cite{Fer16} read
\begin{eqnarray}
\Gamma_{\mu}(t)\!&=&\!-\!\int_0^t\!\!ds\mathbb{D}^{Re}(t,s)\!\left(\cos\omega_S(s-t)+m\mu\omega_S\sin\omega_S(s-t)\right)\nonumber\\
\Theta_{\mu}(t)\!&=&\!\mu\!\int_0^t\!\!ds\mathbb{D}^{Re}(t,s)\!\left(2\cos\omega_S(s-t)- m_-\sin\omega_S(s-t)\right)\nonumber\\
\label{this}\Upsilon_{\mu}(t)&=&-im_+\mu\int_0^tds\,\mathbb{D}^{Im}(t,s)\sin\omega_S(s-t)\,,
\end{eqnarray}
where we have introduced $m_{\pm}=1/m\mu\omega_S\pm m\mu\omega_S$. The integral kernels $\mathbb{D}$ have a series structure that depend both on the bath correlation function and on the free propagator of the system. For the numerical purposes of this paper, the structure of $\mathbb{D}$ represents a drawback because one needs to truncate the series introducing systematic errors. Accordingly, the Heisenberg approach exploited in the main text is more suitable.
On the other hand, Eqs.~\eqref{this} allow to calculate the determinant of $a(t)$ of Eq.~\eqref{kossa} in an easier way. Indeed, we consider the definition of the determinant of $a(t)$
\begin{equation}
\mathrm{det}[a(t)]=4\Gamma_{\mu}(t)\gamma_{\mu}(t)-\left(\Theta_{\mu}(t)^2+\Upsilon_{\mu}(t)^2\right)\,,
\end{equation}
and we replace Eqs.~\eqref{this} in it. By exploiting the composition properties of trigonometric functions, after some calculations we obtain
\begin{eqnarray}
\mathrm{det}[a(t)]&=&-m_+^2\mu^2\left[\left(\int_0^tds\mathbb{D}^{Re}(t,s)\sin\omega_S(s-t)\right)^2\right.\nonumber\\
&&\left.+\left(\int_0^tds\mathbb{D}^{Im}(t,s)\sin\omega_S(s-t)\right)^2\right]\,.
\end{eqnarray}
We then invert Eqs.~\eqref{this} and replace the result in the equation above to eventually obtain Eq.~\eqref{deta}. One can easily check that when the bath is delta correlated the determinant of $a$ is zero.


\begin{thebibliography}{99}

\bibitem{BrePet02} H.P. Breuer and F. Petruccione, {\it Theory of open quantum systems} (Oxford, Oxford University Press, 2002).

\bibitem{Wei08} U. Weiss, {\it Quantum Dissipative Systems} (World Scientific, 2008).

\bibitem{Rivasetal14} {\'A}. Rivas, S.F. Huelga and M.B. Plenio, Rep. Prog. Phys. {\bf 77}, 094001 (2014).

\bibitem{Breetal16} H.-P. Breuer, E.-M. Laine, J. Piilo, and B. Vacchini, Rev. Mod. Phys. {\bf 88}, 021002 (2016).

\bibitem{DeVAlo17} I. de Vega, D. Alonso, Rev. Mod. Phys. {\bf 89}, 015001 (2017).

\bibitem{CalLeg83} A.O. Caldeira, A. Leggett, Physica A {\bf 121}, 587 (1983).

\bibitem{Foretal88} G. W. Ford, J. T. Lewis, R. F. O'Connell, Phys. Rev. A {\bf 37}, 4419 (1988).

\bibitem{HPZ} B.L. Hu, J. P. Paz, Y. Zhang, Phys Rev. D {\bf 45}, 2843 (1992).

\bibitem{HalYu96} J. J. Halliwell, T. Yu, Phys. Rev. D {\bf 53}, 2012 (1996).

\bibitem{ForOco01} G. W. Ford, R. F. O'Connell, Phys. Rev. D {\bf 64},105020 (2001).

\bibitem{Fer16} L. Ferialdi, Phys. Rev. Lett. {\bf 116}, 120402 (2016).

\bibitem{Fer17} L. Ferialdi, Phys. Rev. A {\bf 95}, 020101(R) (2017).

\bibitem{Hall2014} M.J. W. Hall, J.D. Cresser, L. Li, and E. Andersson, Phys. Rev. A {\bf 89}, 042120 (2014).

\bibitem{SanScu87} A. Sandulescu, H. Scutaru, Ann. Phys. {\bf 173}, 277 (1987); A. Isar, A. Sandulescu, H. Scutaru, E. Stefanescu, W. Scheid, Int. J. Mod. Phys. E {\bf 03}, 635 (1994).

\bibitem{Gaoeco} S. Gao, Phys. Rev. Lett. {\bf 79}, 3101 (1997); G. W. Ford, R. F. O'Connell, Phys. Rev. Lett. {\bf 82}, 3376 (1999); S. Gao, Phys. Rev. Lett. {\bf 82}, 3377 (1999).

\bibitem{BasIppVac05} A. Bassi, E. Ippoliti, B. Vacchini,  J. Phys. A {\bf 38}, 8017 (2005).

\bibitem{FerBas12} L. Ferialdi, A. Bassi, Phys. Rev. Lett. {\bf 108}, 170404 (2012); Phys. Rev. A {\bf 86}, 022108 (2012); Europhys. Lett. {\bf 98} 30009 (2012).

\bibitem{Leggett1984} A.J. Leggett, Phys. Rev. B {\bf 30}, 1208 (1984).

\bibitem{Cuccolietal2001} A. Cuccoli, A. Fubini, V. Tognetti, and R. Vaia, Phys. Rev. E {\bf 64}, 066124 (2001)

\bibitem{Kohleretal2004} H. Kohler, F. Guinea, and F. Sols, Ann. Phys. {\bf 310}, 127 (2004);
H. Kohler and F. Sols, New. J. Phys. {\bf 8}, 149 (2006)

\bibitem{DioFer14} L. Di\'osi, L. Ferialdi, Phys. Rev. Lett. {\bf 113}, 200403 (2014).

\bibitem{CarBas16} M. Carlesso, A. Bassi, Phys. Rev. A {\bf 95}, 052119 (2017).

\bibitem{Gorini1976} V. Gorini, A. Kossakowski, and E.C.G. Sudarshan, J. Math. Phys. {\bf{17}}, 821 (1976).

\bibitem{Bengtsson2006} I. Bengtsson and K. Zyczkowski, {\it Geometry of Quantum States: An Introduction to Quantum Entanglement} (Cambridge University Press, Cambridge, UK, 2006).

\bibitem{Lind1976} G. Lindblad, Comm. Math. Phys. {\bf{48}}, 119 (1976).

\bibitem{JoosZeh} E. Joos, H. D. Zeh, Z. Phys. B {\bf 59}, 223 (1985).

\bibitem{Fer17b} L. Ferialdi, Phys. Rev. A {\bf 95}, 052109 (2017). 
 
%\bibitem{Li2017} L. Li and M.J.W. Hall, arXiv:1701.01292 (2017).

\bibitem{Rivas2010} {\'A}. Rivas, S. F. Huelga, and M. B. Plenio, Phys. Rev. Lett. {\bf 105}, 050403 (2010).

\bibitem{Laine2010} E.M. Laine, J. Piilo, and H.-P. Breuer, Phys. Rev. A {\bf 81}, 062115 (2010).

\bibitem{Rivas2012} {\'A}. Rivas and S. F. Huelga, {\it Open Quantum Systems} (Springer, 2012).

\bibitem{Holevo1997} A.S. Holevo, in: A. Bohm, H.D. Doebner and P. Kielanowski (eds) \emph{Irreversibility and Causality}, Lecture Notes in Physics, {\bf 504}, 67 (1997).

%\bibitem{Isar1994} A. Isar, A. Sandulescu, H. Scutaru, E. Stefanescu, W. Scheid, Int. J. Mod. Phys. E {\bf 3}, 635 (1994)

\bibitem{Torre2015} G. Torre, W. Roga, and F. Illuminati, Phys. Rev. Lett. {\bf 115}, 070401 (2015).

\bibitem{Liuzzo2017} P. Liuzzo-Scorpo, W. Roga, L.A.M. Souza, N.K. Bernardes, and G. Adesso, Phys. Rev. Lett. {\bf 118}, 050401 (2017).

\bibitem{Vasile2011} R. Vasile, S. Maniscalco, M.G.A. Paris, H.-P. Breuer, J. Piilo, Phys. Rev. A {\bf 84}, 052118 (2011).

\bibitem{35} A. Smirne and B. Vacchini, Phys. Rev. A {\bf 82}, 042111 (2010).

\bibitem{36} L.A.M. Souza, H. S. Dhar, M.N. Bera, P. Liuzzo-Scorpo, G. Adesso, Phys. Rev. A {\bf 92}, 052122 (2015).


\end{thebibliography}
\end{document}